\documentclass[a4paper,11pt]{article}
\pdfoutput=1 

\usepackage{jcappub} 

\usepackage[T1]{fontenc} 
\usepackage{amsmath, hyperref}
\usepackage{amssymb}
\usepackage{amsfonts}
\usepackage{mathrsfs}
\usepackage{graphicx}
\usepackage{enumerate}

\usepackage{lineno,hyperref}
\modulolinenumbers[5]

\newcommand{\exps}[1]{{\langle #1 \rangle}}

\title{Quantum fluctuations and semiclassicality in an inflaton-driven evolution}

\author[a]{David Brizuela}
\author[b]{Tomasz Paw{\l}owski}

\affiliation[a]{Department of Physics and EHU Quantum Center, University of the Basque Country UPV/EHU, Barrio Sarriena s/n, 48940 Leioa, Spain}
\affiliation[b]{Institute for Theoretical Physics, Faculty of Physics and Astronomy, University of Wroc{\l}aw, pl. M. Borna 9, 50-204  Wroc{\l}aw, Poland}

  \emailAdd{david.brizuela@ehu.eus}
\emailAdd{tomasz.pawlowski@uwr.edu.pl}

\abstract{A semiclassical description of quantum systems is applied to probe the dynamics of the cosmological model of an inflationary
  universe with quadratic inflaton potential, described in a quantum framework of geometrodynamics. The systematic analysis,
  focusing in particular on the inflationary and post-inflationary epochs, revealed several surprising and counterintuitive features: $(i)$ during inflation the universe rapidly spreads out in volume which leads to significant relative variance by the end of inflation; $(ii)$ despite that, the quantum evolution can still be described to high accuracy by semiclassical methods; $(iii)$ moreover, in the post-inflationary epoch, as the order of included quantum corrections increases, the quantum trajectory approaches the classical one and the description involving second-order corrections only is actually the least accurate there. The consequence of the latter is that
  the effects of the quantum variances are washed out by the higher-order quantum corrections.}

\begin{document}
\maketitle
\flushbottom

\section{Introduction}
The $\Lambda$-CDM model is so far commonly accepted as the simplest model of the observed universe explaining the astrophysical observational results (rotation curves of galaxies, cosmological acceleration from supernovae observations \cite{WMAP10, Copeland:2006wr, Huterer17})
with an acceptable level of accuracy. On the other hand, the inflationary paradigm \cite{Sta80, Guth80, Linde81} is another simple model that properly explains
the results of precise measurements in cosmology as, for example, the scale invariance of the cosmic microwave background
power spectra \cite{Planck18}. The value of the cosmological constant determined by the observations
is however extremely small in natural units (about $10^{-120}{\ell}_{\rm Pl}^{-2}$), which leads to the so-called hierarchy problem:
that value is many levels of magnitude smaller than other fundamental constants of nature. Attempts to explain its emergence via field theory as vacuum energy have not been successful
as they predict that it has a very high magnitude (Planckian or trans-Planckian), which led to the so-called
\emph{cosmological constant problem} \cite{Weinberg88, Carroll91, Sahni99, Martin12}.
Hence, the task of providing a viable mechanism of emergence of a cosmological constant of sufficiently small value has gathered much attention in the cosmology community and several approaches have been proposed
(see, for example, \cite{Cohen:1998zx, KQS10, Bull:2015stt, Luongo:2018lgy, Cree18, Carlip18}).

One of the promising approaches is to consider the inclusion of the effects of quantumness of the geometry itself. It is well known
that the basic principles of general relativity and quantum physics are mutually exclusive.
Therefore, these main pillars of modern physics, while very successful, describe the reality accurately in
separate domains (large scales and strong gravitational fields versus small scales and weak gravitational fields).
Yet it is expected that there are domains, where both relativity and quantum effects are important, one of them 
being the early universe. Thus it is crucial to have a consistent framework, not only for philosophical or 
aesthetic reasons, but also in order to construct a self-contained and accurate description of the evolving 
universe.

Representing the (originally classical) spacetime by 
appropriate semiclassical states of very small fluctuations has a priori a potential for generating the (fluctuation-dependent) dynamical effect of the order of magnitude similar to the observed value of the cosmological constant.
A method well suited to analyze such scenario was presented in \cite{Boj05},
and has already successfully been used to analyze the dynamics of several cosmological models;
see, e.g., \cite{ABU21, BU22, Bojowald:2020emy, Baytas:2018gbu, Baytas:2016cbs, Bojowald:2015fla, Brizuela:2015cna}
for recent studies and \cite{Bojowald:2012xy} for a review of previous works.
This method is based on the so-called Hamburger decomposition, where the quantum state is encoded into an infinite set of central moments (functions of expectation values of appropriately chosen observables) and its dynamics as a set of equations of motion for these moments.
It naturally encompasses the quantum corrections up to arbitrary order, allowing in particular to restrict that order by a well defined cutoff of the (originally
infinite) set of equations of motion to a finite one.

In the present work we apply this formalism to study a simple inflationary model given by an isotropic Friedman-Lemaitre-Robertson-Walker universe with a massive scalar field, usually known as the inflaton, with a quadratic potential. While there are much more advanced and accurate (with respect to observations) models, this ``textbook'' model has been studied extensively in the literature, and thus provides an excellent testbed for probing the functionality of the applied techniques and clearly distinguish the effects of higher-order quantum corrections.

In the standard analysis of this physical scenario, one usually considers
the approximation of quantum field theory on curved backgrounds:
the degrees of freedom are split between the background spacetime
and the linearized perturbations of the different fields and, whereas
the quantum behavior of the latter is considered, the background
is approximated by its classical evolution. But if one believes that
the nature is eminently quantum, then a natural question arises:
what happens with the quantum fluctuations of the background objects?
Does the dynamics make them negligible? As there is a classical attractor
for inflationary dynamics, is there a similar (generalized) attractor for the
quantum state of the universe? Does this lead to a classicalization of
the mentioned degrees of freedom as with decoherence
\cite{PoSt95, Giulini96, Kubotani97, Kiefer07},
or as shown in the context of the no boundary proposal \cite{Leh15}?

As a first approximation to the model, we will not consider any inhomogeneous perturbations, but we will study the quantum fluctuations of the background objects and the backreaction they produce on the evolution of a homogeneous cosmological model to a high-order in quantum moments.
This analysis is very relevant since in particular,
as will be explained below, the quantum backreaction will change the classical
trajectory of the universe and, during a long evolution, might produce
effects of the order of the classical perturbations.
Both the matter field and the geometry will be quantized using canonical quantum mechanical tools
(the Wheeler-DeWitt geometrodynamical framework). The goals of our study are the following: $(i)$ test of the technical viability of probing the dynamics of physically interesting quantum cosmological models with high-order quantum corrections via currently available numerical tools; $(ii)$ probing the dynamical effects of high-order corrections during and after inflation (the reheating epoch); and, in particular, $(iii)$ checking whether quantum corrections in the post-inflationary phase can mimic the effect of a small cosmological constant. 

The paper is organized as follows. In Sec. \ref{sec:model} we present the model we will
use to study the corrections of interest and, in particular, the quantum version of the Friedmann
and the Klein-Gordon
equations will be presented. In Sec. \ref{sec:attractors} the stationary attractors of the model are considered
by performing an analytic analysis. Sec. \ref{sec:numerics} deals with the numerical resolution of the dynamics.
Finally, the conclusions are presented in Sec. \ref{sec:conclusions}.

\section{The quantum inflationary model.}\label{sec:model}

The specific model that will be analyzed is an inflaton field $\phi$
with mass $m$ weakly coupled to a homogeneous and isotropic metric
with flat spatial slices. In addition to this matter field, an irrotational dust field $t$
will also be included with the only purpose to provide a time reference. Therefore,
the dust will not be a physical variable and it will not have
any direct effects on the system.
The action of this system is given by,
\begin{equation}\label{eq:action}
  S\ =\ \frac{1}{16\pi G} \int_{\cal V} d^4x \sqrt{-g}R
    - \frac{1}{2} \int_{\cal V} d^4x \sqrt{-g}M(g^{ab}t_{,a}t_{,b}+1)
    - \frac{1}{2} \int_{\cal V} d^4x \sqrt{-g}(g^{ab}\phi_{,a}\phi_{,b}-m^2\phi^2) \ ,
\end{equation}
with $G$ being the Newton gravitational constant and ${\cal V}$ a finite region of the universe,
sometimes also called the fiducial cell.
As it is well known, the Hamiltonian corresponding to this action is vanishing. However,
under the presence of this type of dust, there exists a natural deparametrization
\cite{HuPa11} one can perform. In this representation, the conjugate momentum to the dust field
$p_t$ defines the physical (nonvanishing) Hamiltonian,
\begin{equation}
  {\cal H}\ :=\ -p_t\ =\ {\cal H}_G + {\cal H}_{\phi} \ ,
\end{equation}
which provides the evolution along the time given by the dust $t$.
In this expression, ${\cal H}_G$ and ${\cal H}_{\phi}$ are, respectively, the gravitational and the scalar
field part of the Hamiltonian constraint following from the action \eqref{eq:action}.
In fact, this time choice corresponds to the
usual cosmic time with the lapse $N=1$. The derivatives with respect to this cosmic
time will be denoted by a dot.

Once the above deparametrization has been performed, the matter
degrees of freedom are described by the scalar field $\phi$ and its conjugate
momentum $p_\phi$. Due to the symmetries, there is only one geometric degree of freedom,
which will be described by the canonical pair $(V,\pi_V)$, where $V:=a^3$ stands for the three-volume
of the spatial slice, with $a$ being the scale factor, and, as will be made explicit below,
classically
$\pi_V=-H/(4\pi G)$ is proportional to the Hubble factor $H:=\dot{a}/a$.
The symplectic structure is canonical,
\begin{equation}
  \{V,\pi_V\}\ =\ 1 \ , \qquad \{\phi,p_{\phi}\}\ =\ 1 \ ,
\end{equation}
and, in terms of these variables, the Hamiltonian of the system is explicitly given by,
\begin{equation}\label{classH}
  {\cal H}\ =-6\pi G\, V\pi_V^2+\frac{p_\phi^2}{2 V}+\frac{V}{2}m^2\phi^2.
\end{equation}

The quantization of the system introduces an infinite set of quantum
degrees of freedom, which can be parametrized by the moments,
\begin{equation}\label{moments}
G^{ijkl}:=\langle (\hat{V}-V)^i (\hat{\pi}_V-\pi_V)^j (\hat{\phi}-\phi)^k  (\hat{p}_{\phi}-p_{\phi})^l
	 \rangle_{\rm Weyl},
\end{equation}
where the Weyl subscript stands for totally symmetrical ordering
of the operators, once their powers have been expanded by the usual binomial expansion,
and the expectation values are defined as $V:=\langle \hat V\rangle$,
 $\pi_V:=\langle \hat \pi_V\rangle$, $\phi:=\langle \hat\phi\rangle$,
 and  $p_\phi:=\langle \hat p_\phi\rangle$. The sum of the superindices
 $i+j+k+l$ will be referred as the order of the moment and will
 be used as a guide to truncate the infinite tower of moments.
The evolution of these variables
is given by the expectation value of the Hamiltonian
operator $\hat{\cal H}$. By performing a formal series expansion
around the expectation values, this effective quantum Hamiltonian
takes the form,
\begin{equation}\label{eq:ham-exp}
  {\cal H}_Q\ :=\ \exps{\hat{{\cal H}}(\hat V, \hat{\pi}_V,\hat\phi,\hat p_\phi)}\ =\ {\cal H}(V, \pi_V,\phi,p_\phi) + \sum_{a,b,c,d=0}^{\infty} \frac{1}{a!b!c!d!} \,
    \frac{\partial^{a+b+c+d} {\cal H}}{\partial^a V\, \partial^b\pi_V\, \partial^cp_{\phi}\, \partial^d\phi}
    \, G^{abcd} \ ,
\end{equation}
where we have also assumed a totally
symmetrical ordering of the Hamiltonian operator and ${\cal H}$
is the classical Hamiltonian \eqref{classH}.
As can be seen,
the only term producing terms up to an infinite order in the above expansion is the inverse of volume $V^{-1}$
that appears in the kinetic term of the matter Hamiltonian. Therefore,
one can easily write a more explicit expression for this effective Hamiltonian,
\begin{eqnarray}
 {\cal H}_Q&=&-6\pi G\, V\pi_V^2+\frac{p_\phi^2}{2 V}+\frac{V}{2}m^2\phi^2-6\pi G
 \left(
V G^{0200}+2\pi_V G^{1100}+G^{1200}
 \right)
 + \frac{1}{2V} G^{0002}
 \nonumber\\
 &+&\frac{m^2}{2}\!\left(
V G^{0020}+2\phi G^{1010}+G^{1020}
 \right)\!
 +\!\sum_{n=1}^\infty\frac{ (-1)^n}{2V^{n+1}}\,\left(
G^{n002}+2 p_\phi G^{n001}+p_\phi^2 G^{n000}
 \right)\!.\,\,
\end{eqnarray}
In fact, it turns out to be very useful to rescale the momentum of the field $p_\phi$ as $\pi_{\phi}:=p_\phi/V$ and define the new rescaled quantum moments as,
\begin{equation}
 Q^{ijkl}:=\frac{1}{V^{i+l}}G^{ijkl}.
\end{equation}
In terms of these new variables, the volume appears just as a global factor in the Hamiltonian,
\begin{eqnarray}
 {\cal H}_Q&=&V\Big[-6\pi G\,\pi_V^2+\frac{\pi_{\phi}^2}{2}+\frac{1}{2}m^2\phi^2-6\pi G
 \left(
Q^{0200}+2\pi_V Q^{1100}+Q^{1200}
 \right)
 +\frac{1}{2}Q^{0002}\nonumber\\\label{quantumH}
  &+&\frac{m^2}{2}\left(
Q^{0020}+2\phi Q^{1010}+Q^{1020}
 \right)
 +\!\sum_{n=1}^\infty \frac{(-1)^n}{2}\left(
Q^{n002}+2 \pi_{\phi} Q^{n001}+\pi_{\phi}^2 Q^{n000}
 \right)\!\Big],
\end{eqnarray}
and the evolution equations will take a simpler form, as will be shown below.

In order to obtain the equations of motion, one just needs
to compute the Poisson brackets between different variables and
the Hamiltonian ${\cal H}_Q$ by taking into account that
the Poisson brackets between expectation values are given in terms
of the commutator by the usual relation
$\{\langle \hat A \rangle,\langle \hat B \rangle\}=-\frac{i}{\hbar}\langle[\hat A, \hat B]\rangle$.
In particular, the expectation values of the basic variables commute with the moments,
so it is easy to write their equations of motion explicitly:
\begin{eqnarray}
\dot{V}&=&-12\pi G\, (V \pi_V+ G^{1100}) ,\\
\dot{\pi}_V&=&6\pi G\,\pi_V^2+\frac{p_\phi^2}{2V^2}-\frac{m^2}{2}\phi^2
+6\pi G\, G^{0200}
-\frac{m^2}{2}G^{0020}
+\frac{1}{2V^2}G^{0002}\nonumber\\
&+&\sum_{n=1}^\infty\frac{ (-1)^n}{2V^{n+2}} (n+1)\,\left(G^{n002}+2 p_\phi G^{n001}+p_\phi^2 G^{n000}
 \right),\\
\dot{\phi}&=&\frac{p_\phi}{V}
+\sum_{n=1}^\infty\frac{ (-1)^n}{V^{n+1}}\,\left(
G^{n001}+ p_\phi G^{n000}
 \right),\\
\dot{p}_\phi&=&-m^2V \phi
 - m^2 G^{1010},
\end{eqnarray}
where, as already commented, the dot stands for a derivative with respect to the cosmological time.
The transformation to the rescaled variables remove all explicit volumes,
except from the first equation,
\begin{eqnarray}\label{dotV}
\dot{V}&=&-12\pi G\,V (\pi_V+ Q^{1100}) ,\\
\dot{\pi}_V&=& 6\pi G\,\pi_V^2+\frac{\pi_{\phi}^2}{2}-\frac{m^2}{2}\phi^2
+6\pi G\, Q^{0200}
-\frac{m^2}{2}Q^{0020}+\frac{1}{2}Q^{0002}
\nonumber\\\label{dotpiV}
&+&\sum_{n=1}^\infty\frac{ (-1)^n}{2} (n+1)\,\left(
Q^{n002}+2 \pi_{\phi} Q^{n001}+\pi_{\phi}^2 Q^{n000}
 \right),\\\label{dotphi}
\dot{\phi}&=&\pi_{\phi}
+\sum_{n=1}^\infty(-1)^n\,\left(Q^{n001}+ \pi_{\phi} Q^{n000}
 \right),\\\label{dotpiphi}
\dot{\pi}_{\phi}&=&-m^2 (\phi+ Q^{1010})+8\,\kappa\,\pi_{\phi}(\pi_V+ Q^{1100}) .
 \end{eqnarray}
It is important to note that
these equations correspond just to the evolution equations
for the expectation values. In combination with the equations of motion for the moments $Q^{ijkl}$,
they are part of an infinite system of highly coupled differential equations.
Though they are very long and complicated, one can obtain these equations algorithmically
just by computing the corresponding Poisson brackets with the Hamiltonian \eqref{quantumH}.
For illustration, in the appendix we present the equations of motion for second-order moments
with a sixth-order truncation.
Therefore, in practice, as we will do in the subsequent numerical analysis,
in order to solve this system one needs to introduce a truncation of the infinite
set of moments. The zeroth-order truncation, $Q^{ijkl}=0$ for all $i,j,k,l$,
corresponds to the classical limit. Higher-order truncations, performed by
dropping all $Q^{ijkl}$ from certain order on, introduce with more and
more precision the quantum backreaction effects.
Another interesting limit is that of quantum field theory on classical backgrounds,
which can be obtained just by removing all quantum moments associated to
the geometric degrees of freedom, i. e., just by imposing $Q^{ijkl}=0$ for all $i+j\neq 0$.

Remarkably, when truncated at second order in moments, the system
is manifestly scale invariant (there is no $V$ in the equations for other variables, as can be seen in the appendix), while for higher-order truncations $V$ appears in the equations of motion for the remaining variables only in the form $(\hbar/V)^n$. This implies, in particular, that the system is well defined when the finite region
${\cal V}$ is expanded to encompass the whole universe, i.~e., $V\rightarrow\infty$. Because in the cosmological context the role of such limit carries some analogy\footnote{More precisely, in order to regulate the infinities produced by integrating densities over the whole homogeneous spatial slices when calculating the action, the momenta and the Hamiltonian, one introduces a finite region of the universe ${\cal V}$ (the fiducial cell) that is constant in comoving coordinates. While its physical size changes following the universe expansion, in the homogeneous cases there is no flux of matter or gravitational radiation across its boundaries. The limit of expanding this cell to encompass the whole spatial slice corresponds to the limit $V\to\infty$ at given time.} with the role of the infrared regulator removal limit in field theory, we will also refer to it here as \emph{the infrared regulator removal limit}.
Furthermore, $\hbar$ appears only in the above mentioned terms and
the limit $\hbar\rightarrow 0$ is thus equivalent to the limit $V\rightarrow\infty$.
Hence, in the infrared regulator removal limit
the set of equations of motion becomes mathematically identical to the system corresponding
to a statistical classical ensemble for the geometry \cite{Bri14}.
In consequence, the regulator removal also removes the genuine quantum geometry effects.
 
In order to finish with the presentation of the model,
let us rewrite the evolution equations \eqref{dotV}--\eqref{dotpiphi} as
the quantum version of the Friedmann and Klein-Gordon equations.
On the one hand, taking into account that $V=a^3$, from the Hamiltonian
\eqref{quantumH} and the evolution equations for the geometric variables
\eqref{dotV}--\eqref{dotpiV}, one can obtain
the quantum Friedmann equations,
\begin{eqnarray}\label{friedmann1}
\left( \frac{\dot{a}}{a}\right)^2&=&\frac{4 \pi  G}{3} \left(m^2 \phi^2+\pi_\phi^2+2 \rho_d+ S_{1}\right),\\\label{friedmann2}
\frac{\ddot{a}}{a}&=&-\frac{4 \pi  G}{3} \left(2 \pi_\phi^2-m^2 \phi^2+\rho_d+S_2\right),
\end{eqnarray}
where $\rho_d:=-{\cal H}_Q/V$ is the energy density of the dust matter field.
For the subsequent numerical analysis, this energy density will be chosen to be vanishing so that the dust field does not interfere with
the inflationary dynamics. The rescaled momentum of the scalar field
$\pi_\phi:=p_\phi/V$ is related to the velocity of the field by,
\begin{equation}
\pi_\phi=\frac{\dot{\phi}-S_4}{S_3},
\end{equation}
with the sources,
\begin{subequations}\label{eq:S34}\begin{align}
  S_3 &:= 1+\sum_{n=2}^\infty (-1)^n Q^{n000},  &
  S_4 &:= \sum_{n=1}^\infty (-1)^n Q^{n001}. \tag{\ref{eq:S34}}
\end{align}\end{subequations}
The different sources $S_I$, with $I=1,2,3,4$, completely encode the quantum backreaction
described by the moments $Q^{ijkl}$ and, as can be seen, the classical equations of motion
are straightforwardly obtained by dropping them.
The sources of the Friedmann equations are explicitly given as follows:
\begin{eqnarray*}
S_{1}&:=&12\pi G \left[(Q^{1100})^2-
Q^{0200}-Q^{1200} 
 \right]
 +m^2\left(
Q^{0020}+2\phi Q^{1010}+Q^{1020} 
 \right)
 +Q^{0002}
 \\
 &+&\sum_{n=1}^\infty (-1)^n\,\left(
Q^{n002}+2 \pi_\phi Q^{n001}+\pi_\phi^2 Q^{n000} 
 \right),\\
 S_2&:=& 6\pi G \left[4(Q^{1100})^2+2Q^{0200}-Q^{1200} \right]
 +\frac{m^2}{2}\left(-2 Q^{0020}+2\phi Q^{1010}+Q^{1020}\right)
 +2Q^{0002}
 \\
 &+&\frac{3}{a^3}\dot{(a^3 Q^{1100})}
+\sum_{n=1}^\infty\frac{ (-1)^n}{2} (3n+4)\,\left(
Q^{n002}+2 \pi_\phi Q^{n001}+\pi_\phi^2 Q^{n000} 
 \right).
\end{eqnarray*}
On the other hand, from \eqref{dotphi}--\eqref{dotpiphi}, one can derive the Klein-Gordon equation,
\begin{equation*}
 \ddot{\phi}+\left[3\frac{\dot{a}}{a}-\frac{\dot{S_3}}{S_3}\right]\dot{\phi}+S_3m^2\phi=
 \frac{S_3}{a^3}\dot{\left(\frac{a^3}{S_3}S_4\right)}-S_3m^2Q^{1010}.
\end{equation*}
In fact, it is very clarifying to rewrite this equation as 
\begin{equation}\label{quantumkleingordon}
 \ddot{\phi}+3\frac{\dot{\widetilde a}}{\widetilde a}
 \dot{\phi}+\widetilde{m}^2\phi=3\frac{\dot{\widetilde a}}{\widetilde a}S_4
 +\dot{S_4}-\widetilde{m}^2Q^{1010},
\end{equation}
with an effective scale factor and (time-dependent) mass defined by
$\widetilde a^3:=a^3/S_3$  and $\widetilde{m}^2:=S_3 m^2$ respectively.
These are indeed the scale factor and the mass effectively seen
by the scalar field, and they are the corresponding classical objects
corrected by the pure fluctuations of the volume $Q^{n000}$
that appear in the source term $S_3$.
Note also that, unlike the classical equation,
the quantum equation \eqref{quantumkleingordon} is not homogeneous
due to the correlations between the matter and the volume
that appear in the right-hand side
and act as a source for the inflaton field.

Finally, it is straightforward to see that, from the Friedmann equations above \eqref{friedmann1}--\eqref{friedmann2}, one might interpret the quantum
backreaction terms as a perfect fluid with effective energy density
$\rho_{\rm eff}:=S_1/2$ and pressure $p_{\rm eff}:=(2 S_2-S_1)/6$.
This interpretation will be useful later to describe the physical
behavior of
the remnant energy of the quantum fluctuations after inflation. 
Nonetheless, one needs to be careful as
both $\rho_{\rm eff}$ and $p_{\rm eff}$ may take any real value
and, in particular, may be negative.

\section{Attractors}\label{sec:attractors}

Even if the system of equations is very involved, there are some physically relevant
conclusions that can be drawn by an analytical analysis. It is well known that the phase-space
point $(\pi_V=0, \phi=0,p_\phi=0)$ is a stationary attractor for the classical system.
For the quantum model, we have performed a stability analysis up to sixth-order in moments
by imposing the vanishing of all time derivatives in a first-order formulation of
the system of equations and solving the remaining algebraic system to find the stationary points.

In particular, considering the truncation at second order in moments (that is, imposing $Q^{ijkl}=0$ for all $i+j+k+l>2$),
there are only three independent exact stationary points that do not require the vanishing of the volume $V$.
On the one hand, one gets a semiclassical solution that generalizes the classical one with
$(\pi_V=0, \phi=0,p_\phi=0)$. This solution is given by the vanishing of all quantum moments, except for the
fluctuation of the matter field $Q^{0020}$ and its conjugate momentum $Q^{0002}$,
which in turn are constant and must obey the relation $Q^{0002}= m^2 Q^{0020}$.
At this solution the Hamiltonian takes the form ${\cal H}_Q=m^2VQ^{0020}$ and it resembles the stationary point of a
harmonic oscillator with frequency $m$ (see, e.g., \cite{Bri14b}).
In fact, at this stationary solution, one recovers the limit
of quantum field theory on curved backgrounds commented above. It is straightforward to see
from the Friedmann equations \eqref{friedmann1}--\eqref{friedmann2} that these quantum fluctuations can be interpreted as an effective dust
field with constant energy density $\rho_{\rm eff}=(m^2Q^{0020}+Q^{0002})/2=m^2 Q^{0020}$,
but not as a cosmological constant.
On the other hand, the other two solutions for the second-order truncation in moments
are more complicated and can not be considered semiclassical since 
some of the quantum moments are required to be of the order of the
classical quantities.

Furthermore, all the commented solutions violate the uncertainty relation for
the geometric degrees of freedom $(V,\pi_V)$. However, interestingly, the semiclassical solution
obeys the uncertainty relation for the field variables $(\phi,\pi_\phi)$.
This means that this solution would be dynamically reachable
only if the geometric variables $(V,\pi_V)$ are ``classicalized'' and thus are allowed to
violate the uncertainty relation. In fact, this is somehow what happens
for large volumes since, as explained above, the limit
$V\rightarrow\infty$ is equivalent to $\hbar\rightarrow 0$.

Nonetheless, even if this semiclassical solution might be a very nice
generalization of the classical behavior,
the same stationary analysis considering higher-order truncations in moments shows that
this semiclassical stationary point is spoiled by the backreaction generated by
higher-order moments. This is a first indication that, as will be shown in
detail in the numerical analysis, second-order results are very
particular and that generic (high-order) behavior can not be
directly inferred from this order.

\section{Numerical analysis}\label{sec:numerics}

As the main goal of our study is the analysis of the effects of the high-order quantum corrections on the dynamics, we now focus on finding the solutions to the truncated equations of motions at a given order, in particular probing their dependence on this truncation order. Since these equations constitute a large number
of highly coupled nonlinear ordinary differential equations, apart from being extremely difficult to apply, any approximated analytical
analysis would wash out the effects we are trying to capture. We thus have to rely on a numerical analysis, ensuring the robustness of the results by choosing sufficiently large population of evolved cases and sufficiently wide set of initial data. More precisely, we have considered the truncation of the system
at consecutive orders in moments, from the second to the sixth one, which gives a set of finite systems
of rapidly growing size (from $14$ equations at second order, up to $209$ equations at sixth order).

In order to fix the initial state before the onset of inflation,
we have chosen two different shapes of wave functions: a Gaussian in
the volume $V$ and a Gaussian in the logarithm of the volume $\ln V$,
both peaked around their corresponding classical value. 
While the former are the coherent states in the selected quantum variables, the latter are not, while still being sharply peaked and satisfying the requirements of the semiclassical states.
The initial data were set 
at near Planck densities, where the effect of the scalar field potential is small.
The probed population involved about 200 datasets with the initial value of the scalar
field $\phi$ set to ensure a long inflationary epoch. In addition, the initial
ratios between variances in canonical pairs, $G^{0020}/G^{0002}$ for the scalar
field and $G^{2000}/G^{0200}$ for the geometric variables,
were respectively varied from $0.1$ to $10$ and from $0.5$ to $2$. 
Finally, to avoid possible instabilities that could a priori appear due to the
presence of terms quadratic in moments $G^{abcd}$, the initial volume has been set
to be large (in most simulations of the order of $V=10^{12}\ell_{\rm PL}$)
in order to suppress these terms (which are proportional to inverse powers of $V$).
The resulting initial value problems (composed by the truncated equations of motion and the
initial data chosen as described above) have been subsequently integrated via the explicit
adaptive 4-5th order Runge-Kutta method (Cash-Carp). By direct inspection
(checking the convergence of the solutions as the error tolerances are decreased), we have seen
that the equations of motion have been stable within the domain of variables
covered by our choice of the initial data. Thus, despite the presence of quadratic terms,
the choice of an explicit method still yield reliable solutions.

\begin{figure*}[t]
  \includegraphics[width=7.7cm]{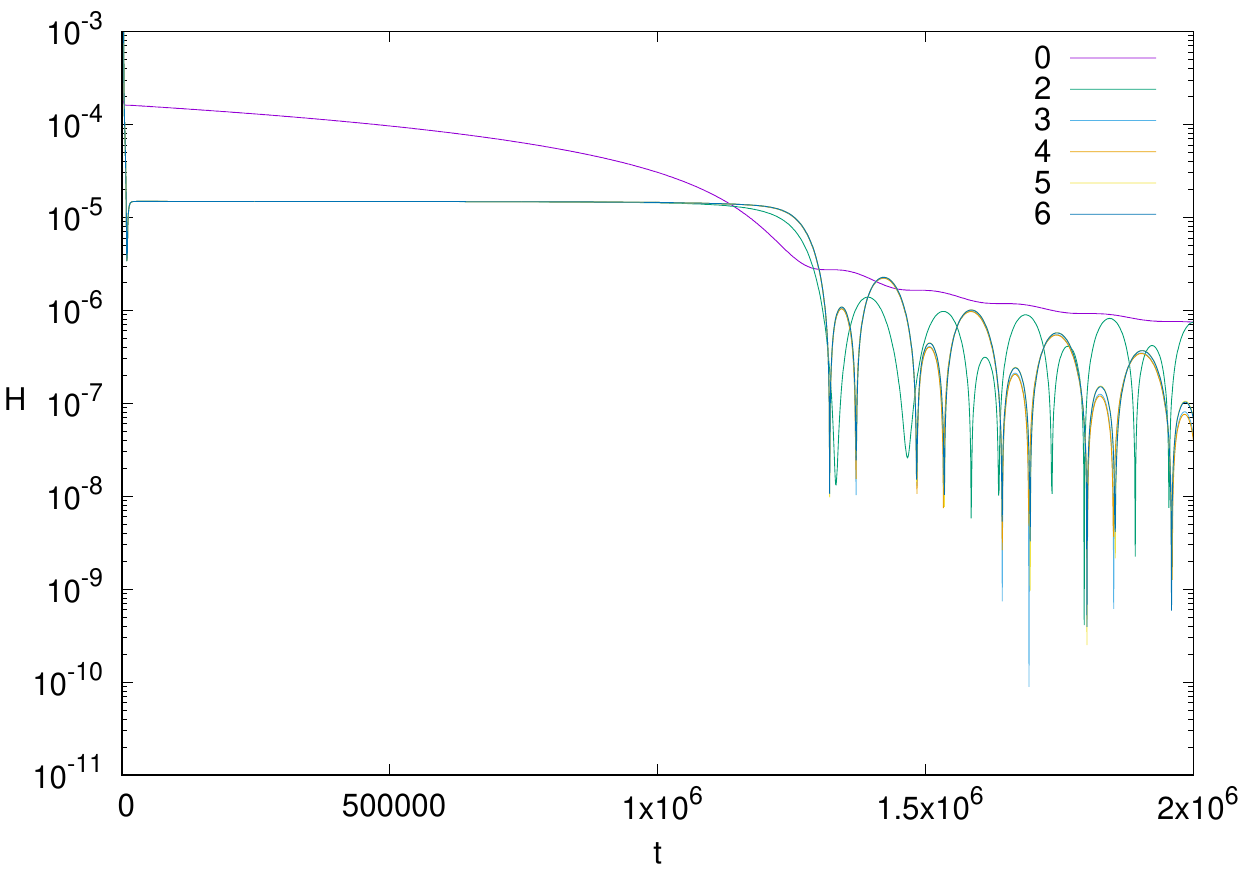}
  \includegraphics[width=7.7cm]{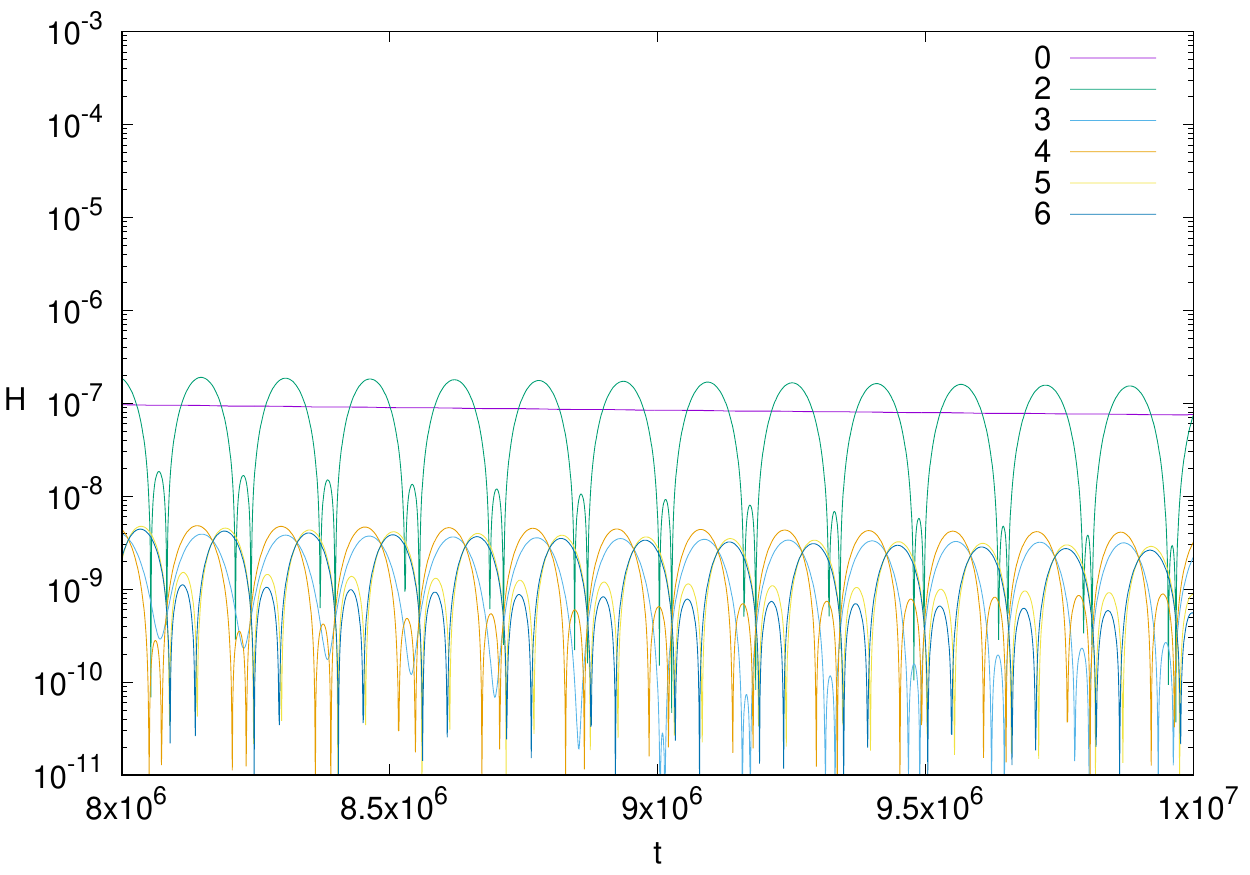}
  \caption{\label{fig:Hr} The purple curve shows the evolution of the classical (0th-order) Hubble parameter
  in a logarithmic scale. Its quantum
  correction (defined as the difference of the Hubble factor at every truncation order and its classical value),
  multiplied by a factor of $10^5$ for presentation, is depicted in different colors denoting the truncation order.
  The plot on the left shows the inflationary period and the beginning of the reheating epoch, while the plot on
  the right corresponds to well after the inflation exit. In this latter plot, one can see the oscillations of the second-order corrections over a non-zero equilibrium point. However, as the truncation order increases,
  both the amplitude and the equilibrium value of the oscillations decrease.}
\end{figure*}

As expected, the classical trajectories of the relevant physical parameters are slightly corrected due to quantum backreaction terms.
However, the nature and relevance of these corrections depends on the epoch of the evolution. During the inflation, the classical trajectories of quantities invariant under volume rescaling
(that is, the field $\phi$, the relative momentum $\pi_\phi$, and the Hubble rate $H$)
are accurately
corrected by the second-order equations of motion, and there is a very fast
convergence of the trajectories as the truncation order increases. The corrections due to considering higher
orders are at least $1-2$ orders of magnitude lower than the corrections corresponding to the second-order itself.
An example of this is shown in Fig.~\ref{fig:Hr}, where the classical Hubble rate $H$ has been depicted,
alongside with its quantum correction (defined as the absolute value of the difference with its classical value) for different truncation orders.
As can be seen, during the inflationary period (plot on the right of Fig.~\ref{fig:Hr}) $H$ slowly decreases its value, while its quantum correction is approximately constant and equal for
all the considered orders. As the system approaches the end of inflation, the quantum correction
begins to oscillate. Note that the plot is logarithmic and the peaks of the curves
are simply points where the corresponding variable vanishes.

Then, upon exit from the inflation the situation changes (plot on the left of Fig.~\ref{fig:Hr}).
At the second-order truncation we observe that there is a slowly decaying 'vacuum energy'
since the quantum correction of the Hubble factor oscillates around a positive
(slowly decaying) value. However,
instead of amplifying this effect, the higher-order corrections counter-balance it and push the trajectory nearer the classical trajectory. The \emph{time-averaged} correction to the Hubble factor decreases by several orders of magnitude as the truncation order is increased and, in addition, it decays faster in time. In consequence, the inclusion of higher-order terms gives
an evolution very close to the classical predictions.

Concerning the evolution of the moments, the actual ratios between different moments do depend on the precise values of the initial data as well as on the shape of the initial state (ordinary versus logarithmic Gaussian).
Therefore, we conclude that there is not an exact global attractor state, even if the qualitative behavior of the different moments appears to be very similar for all tested initial data.

\begin{figure}[t]
\begin{center}
  \includegraphics[width=9cm]{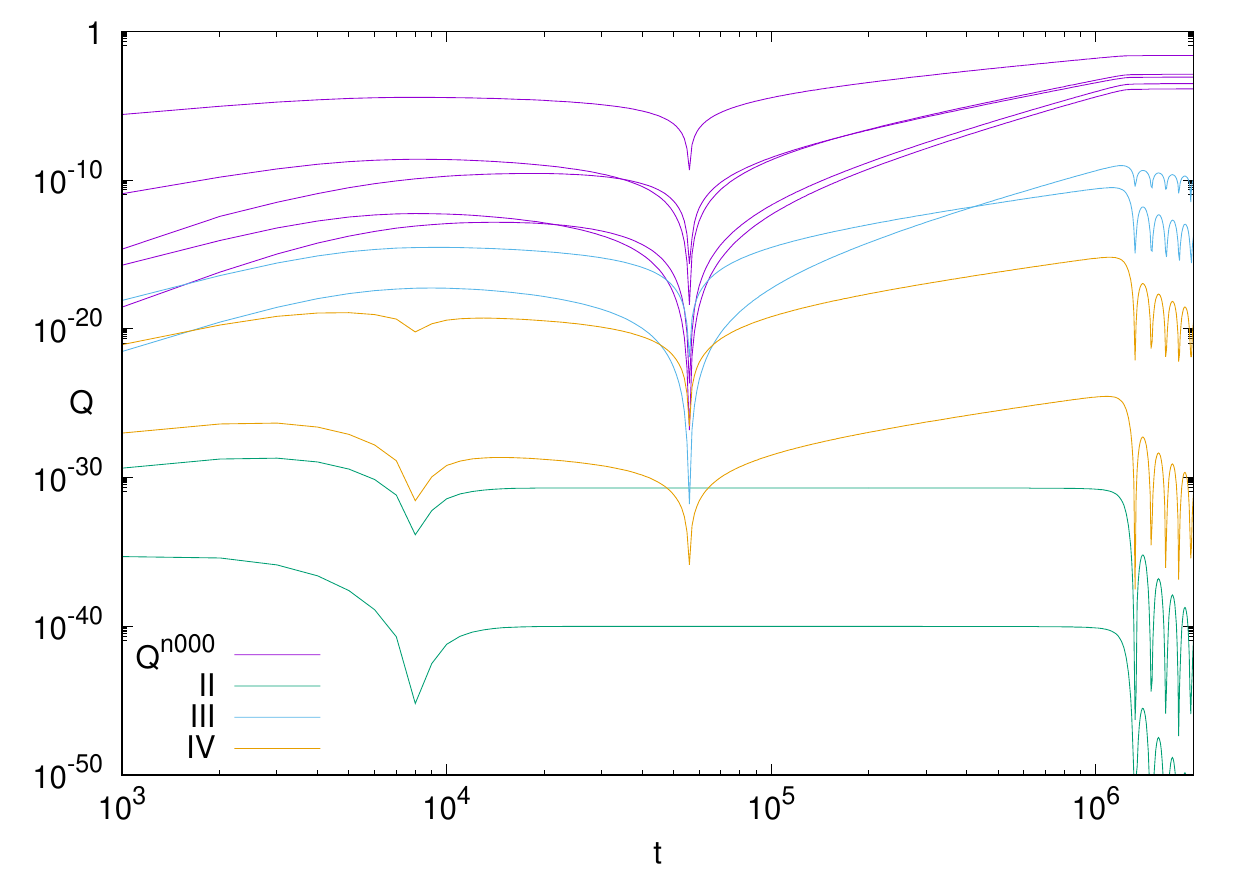}
\caption{\label{fig:Ks} The evolution of the absolute value of the logarithm of the relative central moments $\ln |Q^{ijkl}|$
plotted for several representative examples. The spikes mark the onset of inflation (at $t=t_{\rm on}\approx 8\times 10^4 t_{Pl}$)
and mid-inflation transition (at $t=t_{\rm mid}\approx 6\times 10^5t_{Pl}$), whereas
the oscillations mark the post-inflationary era.
Pure-volume moments $Q^{n000}$ grow during late inflation and stabilize after.
Non-volume moments $Q^{0ijk}$ (group II) stabilize during and decay past inflation. Finally,
mixed moments ($Q^{ijkl}$ with $i\neq 0$ and $j+k+l\neq 0$) grow during late inflation and decay past the inflation exit. These are  denoted by
groups III and IV, which differ by the sign flip of the latter at the onset of inflation.
}
\end{center}
\end{figure}

Remarkably, on top of the onset of inflation (which happens at $t_{\rm on}\approx 8\times 10^4 t_{Pl}$, with $t_{Pl}$ being the Planck time)
and exit from inflation $(t_{\rm exit}\approx 1.2\times 10^6 t_{Pl})$, we observe one more mid-inflation transition point where the behavior of the moments changes.
\footnote{Note that the specific values of the different times we are providing here
correspond to the example shown in Fig. \ref{fig:Ks}.}
This can be seen in the spikes of some
of the curves depicted in Fig. \ref{fig:Ks} around $t_{\rm mid}\approx6\times10^5 t_{Pl}$.
To understand the role of this point (and to exclude the possibility of it being an artifact of numerical instabilities), a more detailed analysis of the second-order system was performed. Since at this order of truncation the system is linear in moments, it was possible to analytically probe the properties
of the subsystem for the moments only with the classical variables treated as ``background''.
That is, writing this subsystem as $\dot{\omega}=A\cdot\omega$, with $\omega$ being the vector column composed by
the second-order moments and the matrix $A$ depending on the classical variables,
one can analyze the eigenvalues of this matrix. It turns out that
the structure of these eigenvalues changes depending on the sign of the function
\begin{equation}
\left(m^2-12 \pi  G \pi_{\phi}^2\right)^2
-48 \pi  G m^2 \phi\left(3 \pi_{\phi} H+m^2 \phi\right),
\end{equation}
which distinguishes the transition point. More precisely, this function
appears in the expression of most of the eigenvalues inside a square root and
switches sign (from negative to positive) at this transition point.
Therefore, the eigenvalues go from being complex to being purely real,
which produces a sudden qualitative change in the evolution of the moments.

For illustration, in respect of the qualitative behavior of the moments, we show some representative examples in Fig.~\ref{fig:Ks}. Their main properties can be summarized as follows:
\begin{itemize}
\item Before inflation $(t<t_{\rm on})$, the majority of the moments increase in absolute value, which corresponds to the spreading-out of the initially coherent state.
\item During early inflation $(t_{\rm on}<t<t_{\rm mid})$, all the moments either remain constant or decrease in absolute value, till the transition discussed in the paragraph above occurs at $t_{\rm mid}$.
\item After the transition $(t_{\rm mid}<t<t_{\rm exit})$ all the moments, except the moments unrelated to the volume $Q^{0ijk}$, rapidly increase following certain power law.
\item After inflation ends $(t_{\rm exit}<t)$, the \emph{volume moments} $Q^{n000}$ become constant,
while all the remaining ones rapidly decrease in absolute value as they oscillate.
\end{itemize}
Note that, taking these properties into account, in Fig.~\ref{fig:Ks} four different
group of moments have been distinguished. On the one hand, volume moments $Q^{n000}$
form the first group, while
moments unrelated to volume $Q^{0ijk}$ are denoted as group II. On the other hand,
mixed moments, i.e., $Q^{ijkl}$ with $i\neq 0$ and $j+k+l\neq 0$, form both groups
III and IV. The difference between these two groups is that, whereas moments
of group IV flip sign at the onset of inflation, moments of group III do not change
sign until the transition point at $t_{\rm mid}$.

All in all,
one can state that inflation
causes rapid dispersion (spreading out) of the state in the volume direction:
while the moments in $\pi_V$ are kept constant
(or suppressed), the volume moments $Q^{n000}$ are amplified by several orders of magnitude, and stabilize only after exiting inflation.
Furthermore, consecutive pairs of higher volume
moments are suppressed just by about an order of magnitude per pair,
that is, $Q^{n000}/Q^{(n+2)000}\approx 10$.

Since the volume moments $Q^{n000}$ are clearly dominating and approximately
constant at the end of inflation, one can compute the sources of the
Friedmann \eqref{friedmann1}--\eqref{friedmann2} and Klein-Gordon \eqref{quantumkleingordon}
equations at that point. In particular one can see
that all the sources are given in terms of the positive sum
$Q:=\sum_{n=2}^\infty(-1)^nQ^{n000}$ as $S_1=2 S_2=\pi_\phi^2 Q$, $S_3=1+Q$, and $S_4= 0$. 
Therefore, the remnant quantum contributions at the end of inflation
to the Friedmann equations can be interpreted as a perfect fluid
with vanishing pressure $p_{\rm eff}=0$ and positive effective energy
density $\rho_{\rm eff}=\pi_\phi^2 Q/2$. This equation of state
corresponds to a dust field, and not to a cosmological constant.
Concerning the evolution of the scalar field, the right-hand side
of the Klein-Gordon equation \eqref{quantumkleingordon} vanishes.
Thus, during reheating, the scalar field simply evolves with an effective scale factor
and mass respectively defined by $\widetilde{a}^3=a^3/S_3$ and $\widetilde{m}^2=S_3 m^2$.

\section{Conclusions}\label{sec:conclusions}

In this paper we have analyzed the effects of the quantum backreaction during and just after
the inflationary period of the universe. For such a purpose, we have considered
the simple model of a weakly coupled inflaton field with mass $m$.
The quantum version of the Friedmann \eqref{friedmann1}--\eqref{friedmann2}
and Klein-Gordon \eqref{quantumkleingordon} equations have been obtained, where
the quantum backreaction is described by the presence of the source terms $S_I$,
which are explicitly given in terms of quantum moments.
After analyzing this system with both analytical and numerical methods,
our main results can be summarized as follows.

On the one hand, concerning the {\textit{quantumness}} of the geometry, 
the form in which the volume $V$ appears in the equations of motion
forces the genuine quantum geometry effects to be washed out in the infrared regulator removal. That is, the limits $V\rightarrow\infty$ and $\hbar\rightarrow 0$ are equivalent.
Thus, in this limit, the system becomes indistinguishable from a classical statistical ensemble.

On the other hand, regarding the dynamical evolution of the system, our numerical studies have revealed the following properties:

\indent{$\bullet$} While there does not seem to exist an exact 
    attractor state, for all the analyzed cases, inflation amplifies 
    the relative fluctuations of the volume while diminishing
    the relative fluctuations of the Hubble factor.
    
\indent{$\bullet$} The quantum remnants by the end of inflation can be
effectively described by a perfect fluid with a vanishing pressure, which corresponds
to a dust field but not to a cosmological constant.
    
\indent{$\bullet$} While during inflation the quantum corrections are accurately captured already at the second-order truncation,
in the post-inflationary epoch the second-order corrections are balanced with the truncation order. As this order is increased,
the trajectories approach the classical one and thus the second-order truncation provides the least accurate results.

The latter is particularly important when considering the accuracy of the classical effective approximations and second-order corrections in cosmology, as getting reliable results will often require verification by including higher-order corrections.
It is important to remember that the simple model used in our studies is essentially a textbook example. It however offered an excellent testing ground, thus opening the possibility of performing similar studies also in more advanced and accurate cosmological models. Furthermore, the general conclusions drawn from our studies are expected to hold to a significant degree also in these more advanced models.
 
While some of the above results have been obtained by means of numerical methods
(and thus, by necessity, the studies were restricted to a finite number of data examples), their population was large and the range of data sufficiently wide to establish the robustness of the results. It has to be however remembered, that the analysis still has its limitations. For example, the initial variances have been set so that their order of magnitude (in Planck units) between
conjugate variables differs at most by a level of magnitude. While within the probed domain the features reported above were not sensitive to the initial data choice, one cannot exclude differences in behavior once states
that are extremely squeezed already near the initial singularity are considered.

Finally, let us note that in our study we have computationally implemented the conceptually uncomplicated
technique of moment decomposition of the wave function. Even if the theoretical basis of this approach
has been widely known in quantum optics for long time,
thanks to the progress in computing devices, now it can be systematically applied to a wide range of
quantum-mechanical systems of interest, not only restricted to cosmology.

\acknowledgments
We acknowledge financial support from the Basque Government Grant \mbox{No.~IT1628-22},
from the Grant FIS2017-85076-P, funded by
MCIN/AEI/10.13039/501100011033 and by “ERDF A way of making Europe”,
and from the Polish Narodowe Centrum Nauki (NCN)
grants 2012/05/E/ST2/03308 and 2020/37/B/ST2/03604.

\appendix

\section{Equations of motion for second-order moments}\label{sec_appendix}

The equations of motion for second-order moments within a sixth-order truncation,
that is, neglecting all moments $Q^{ijkl}$ with $i+j+k+l>6$, read as follows,
\begin{eqnarray*}
 \dot{Q}^{2000}&=&Q^{1100} (24\pi G\,  Q^{2000}-24\pi G\,
   )-24\pi G\,  Q^{2100},
   \\ \dot{Q}^{0200}&=&-2 \phi  Q^{0110}
   m^2-Q^{0120} m^2+2 \pi_{\phi}  Q^{0101}+Q^{0102}+24\pi G\,  \pi_V Q^{0200}+12\pi G\,  Q^{0300}
   \\
   &-&2 \pi_{\phi}
   ^2 Q^{1100}-4 \pi_{\phi}  Q^{1101}-2 Q^{1102}+3
   \pi_{\phi} ^2 Q^{2100}+6 \pi_{\phi}  Q^{2101}+3
   Q^{2102}-4 \pi_{\phi} ^2 Q^{3100}\\&-&8 \pi_{\phi}
   Q^{3101}-4 Q^{3102}+5 \pi_{\phi} ^2 Q^{4100}+10 \pi_{\phi}
    Q^{4101}-6 \pi_{\phi} ^2 Q^{5100},
    \\ \dot{Q}^{0020}&=&2
   Q^{0011}-2 \pi_{\phi}  Q^{1010}-2 Q^{1011}+2 \pi_{\phi}
    Q^{2010}+2 Q^{2011}-2 \pi_{\phi}  Q^{3010}-2
   Q^{3011}+2 \pi_{\phi}  Q^{4010}\\&+&2 Q^{4011}-2 \pi_{\phi}
    Q^{5010},
    \\ \dot{Q}^{0002}&=&-2 Q^{0011} m^2-2 \phi
   Q^{1001} m^2-2 Q^{1011} m^2+Q^{0002} (24\pi G\,
   \pi_V+24\pi G\,  Q^{1100}),
         \end{eqnarray*}
   \begin{eqnarray*}
   \dot{Q}^{1100}&=&-\phi
   Q^{1010} m^2-\frac{1}{2} Q^{1020} m^2+12\pi G\,
   (Q^{1100})^2-12\pi G\,  Q^{0200}+\pi_{\phi}
   Q^{1001}+\frac{1}{2} Q^{1002}\\&+&12\pi G\,  \pi_V
   Q^{1100}-6\pi G\,  Q^{1200}-\pi_{\phi} ^2 Q^{2000}-2
   \pi_{\phi}  Q^{2001}-Q^{2002}+\frac{3}{2} \pi_{\phi} ^2
   Q^{3000}+3 \pi_{\phi}  Q^{3001}\\&+&\frac{3}{2} Q^{3002}-2
   \pi_{\phi} ^2 Q^{4000}-4 \pi_{\phi}  Q^{4001}-2
   Q^{4002}+\frac{5}{2} \pi_{\phi} ^2 Q^{5000}+5 \pi_{\phi}
   Q^{5001}-3 \pi_{\phi} ^2 Q^{6000},
   \\ \dot{Q}^{1010}&=&-12\pi G\,
   Q^{0110}+Q^{1001}+12\pi G\,  Q^{1010} Q^{1100}-12\pi G\,  Q^{1110}-\pi_{\phi}  Q^{2000}-Q^{2001}
   \\&+&\pi_{\phi}
    Q^{3000}+Q^{3001}-\pi_{\phi}
   Q^{4000}-Q^{4001}+\pi_{\phi}
   Q^{5000}+Q^{5001}-\pi_{\phi}
   Q^{6000},\\ \dot{Q}^{1001}&=&-Q^{1010} m^2-\phi
   Q^{2000} m^2-Q^{2010} m^2-12\pi G\,
   Q^{0101}+Q^{1001} (12\pi G\,  \pi_V+24\pi G\,
   Q^{1100})\\&-&12\pi G\,  Q^{1101},
   \\ \dot{Q}^{0110}&=&-\phi
   Q^{0020} m^2-\frac{1}{2} Q^{0030} m^2+\pi_{\phi}
   Q^{0011}+\frac{1}{2} Q^{0012}+Q^{0101}+12\pi G\,
   \pi_V Q^{0110}+6\pi G\,  Q^{0210}\\&-&\pi_{\phi} ^2
   Q^{1010}-2 \pi_{\phi}  Q^{1011}-Q^{1012}-\pi_{\phi}
   Q^{1100}-Q^{1101}+\frac{3}{2} \pi_{\phi} ^2 Q^{2010}+3
   \pi_{\phi}  Q^{2011}+\frac{3}{2} Q^{2012}
   \\
   &+&\pi_{\phi}
   Q^{2100}+Q^{2101}-2 \pi_{\phi} ^2 Q^{3010}-4 \pi_{\phi}
    Q^{3011}-2 Q^{3012}-\pi_{\phi}
   Q^{3100}-Q^{3101}+\frac{5}{2} \pi_{\phi} ^2 Q^{4010}
   \\&+&5
   \pi_{\phi}  Q^{4011}+\pi_{\phi}  Q^{4100}+Q^{4101}-3
   \pi_{\phi} ^2 Q^{5010}-\pi_{\phi}
   Q^{5100},
   \\ \dot{Q}^{0101}&=&-\phi  Q^{0011} m^2-\frac{1}{2}
   Q^{0021} m^2-Q^{0110} m^2-\phi  Q^{1100}
   m^2-Q^{1110} m^2+\pi_{\phi}  Q^{0002}+\frac{1}{2}
   Q^{0003}\\&+&6\pi G\,  Q^{0201}-\pi_{\phi} ^2 Q^{1001}-2
   \pi_{\phi}  Q^{1002}-Q^{1003}+Q^{0101} (24\pi G\,
   \pi_V+12\pi G\,  Q^{1100})\\&+&\frac{3}{2} \pi_{\phi} ^2
   Q^{2001}+3 \pi_{\phi}  Q^{2002}+\frac{3}{2} Q^{2003}-2
   \pi_{\phi} ^2 Q^{3001}-4 \pi_{\phi}  Q^{3002}-2
   Q^{3003}+\frac{5}{2} \pi_{\phi} ^2 Q^{4001}\\&+&5 \pi_{\phi}
   Q^{4002}-3 \pi_{\phi} ^2
   Q^{5001},\\ \dot{Q}^{0011}&=&-Q^{0020} m^2-\phi
   Q^{1010} m^2-Q^{1020} m^2+Q^{0002}-\pi_{\phi}
   Q^{1001}-Q^{1002}
   +\pi_{\phi}  Q^{2001}+Q^{2002}\\&-&\pi_{\phi}
    Q^{3001}-Q^{3002}+\pi_{\phi}
   Q^{4001}+Q^{4002}-\pi_{\phi}  Q^{5001}
+12\pi G\,  Q^{0011} (\pi_V+ Q^{1100}).
   \end{eqnarray*}

\providecommand{\noopsort}[1]{}\providecommand{\singleletter}[1]{#1}%

\end{document}